\RequirePackage{fix-cm}
\documentclass[smallextended]{svjour3}       
\smartqed  

\usepackage{lipsum}
\makeatletter
\def\ps@pprintTitle{%
 \let\@oddhead\@empty
 \let\@evenhead\@empty
 \def\@oddfoot{}%
 \let\@evenfoot\@oddfoot}
\makeatother

\usepackage{amssymb}
\usepackage[table,xcdraw]{xcolor}

\usepackage{epstopdf} 
\usepackage{array}
\usepackage{color}
\usepackage{tabularx}
\usepackage{graphicx}
\usepackage{amsmath}
\usepackage{amssymb}
\usepackage{amsfonts}
\usepackage{multirow}
\usepackage{environ}
\usepackage{yhmath}
\usepackage[utf8]{inputenc}
\usepackage{pdfpages}
\journalname{Medical \& Biological Engineering \& Computing}
\begin{document}

\title{Parameter identification method for breast biomechanical numerical model
}


\author{Diogo Lopes         \and
        Stéphane Clain \and
        António Ramires Fernandes 
}


\institute{Diogo Lopes \at
              Centro Algoritmi, University of Minho, \\
              Campus of Gualtar, 4710-057 Braga, Portugal \\
              \email{diogo.a.rocha.lopes@gmail.com}           
           \and
           Stéphane Clain \at
              Centre of physics, University of Minho, \\
              Campus of Gualtar, 4710-057 Braga, Portugal \\
              \email{clain@math.uminho.pt}
            \and
            António Ramires Fernandes \at
                Centro Algoritmi, University of Minho, \\
                Campus of Gualtar, 4710-057 Braga, Portugal \\
                \email{arf@di.uminho.pt}
}


\maketitle

\begin{abstract}
An accurate numerical model of the breast can help surgeons making better informed decisions by providing visual information of the final aspect of a breast after a surgery simulation. Surgery simulators can help surgeons decide which path to take during a surgery in order to increase its chance of success.

Bio-mechanical breast simulations are based on a gravity free geometry as a reference domain and a nonlinear mechanical model characterised by physical coefficients. Most of the models proposed in literature are complex and use data from medical imaging to estimate said parameters. In some countries, the access to medical imaging devices is limited, hence, we propose a simple but yet realistic model that uses only a basic set of measurements easy to do in the context of routinely operations to obtain the bio-mechanical parameters of the breast.

Both the mechanical system and the geometry are controlled with parameters we shall identify in an optimisation procedure. We give a detailed presentation of the model together with the optimisation method and the associated discretisation. Sensitivity analysis is then carried out to evaluate the robustness of the method.

\end{abstract}

\section*{Author Biography}

- Diogo Lopes: PhD on Science Computing involving numerical simulations, namely breast reduction surgeries simulation, neural networks and high performance computing. \\
- Stéphane Clain: Professor at Universidade do Minho, He has many published articles on finite element methods and use of numerical methods for simulations and HPC. \\
- António Ramires Fernandes: Professor at Universidade do Minho, He possesses vast experience on computer graphics and science computing involving simulations and HPC.

\section*{Graphical Abstract}
\begin{figure}[ht]
\includegraphics[width=0.95\textwidth]{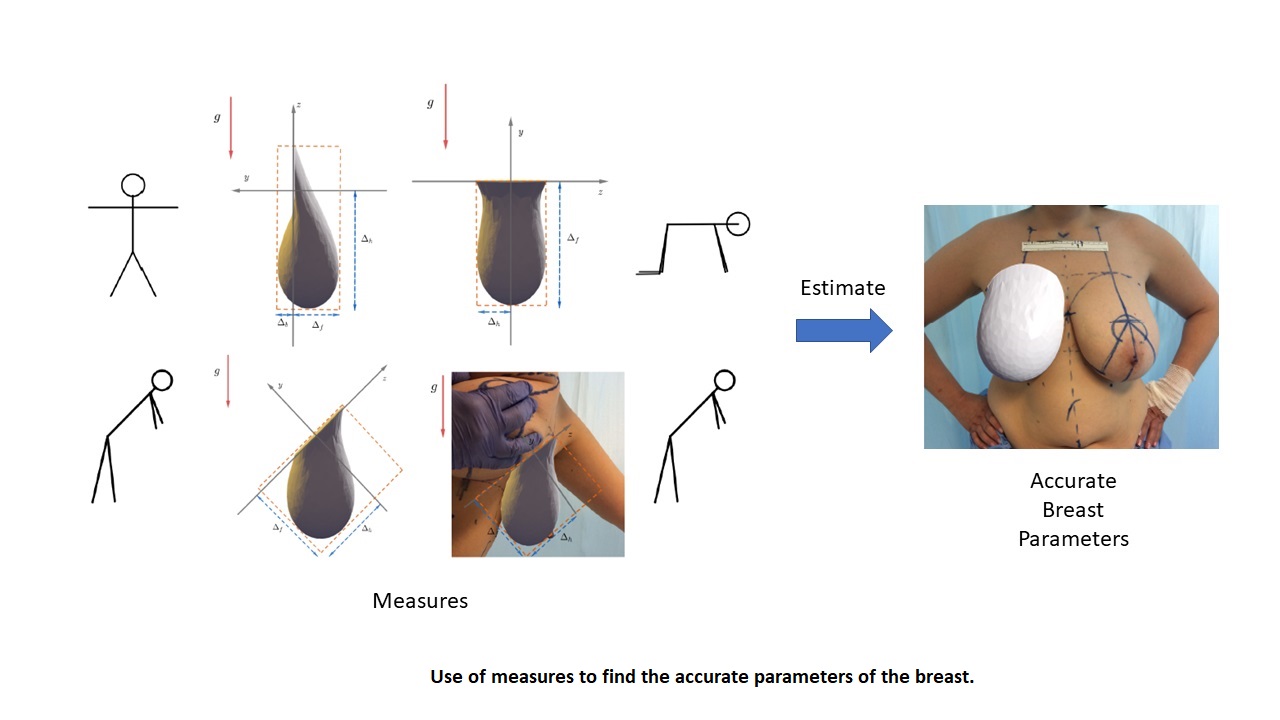}
\centering
\caption{Process of obtaining measurements from patient to obtain breast parameters.}
\label{fig:ga}
\end{figure}

\section{Introduction}

 Breath reduction is a rather routinely operation but requires patient preparation to design the cuts and the nipple new collocation. Numerical simulations of soft tissues turn to be a powerful tool to test surgery scenario and make predictions. The parameters associated to soft material that characterise the breath and the skin are key ingredients in the constitutive model's performance. The correct identification and estimation of the geometric and mechanics parameters is a crucial issue for achieving representative purposes as the study in \cite{ERJ2014} shows. From \textit{ex vivo} tissue samples to elastography there exists a large spectrum of techniques that are used to estimate the elastic properties of the female breast. The identification of soft tissue parameters depends on the underlying model and the material characterisation, like, the linear elastic model, the Neo-Hookean model or even the Mooney-Rivlin model which are expressed in the stress-free unloaded reference state of the breast (\cite{ERJ2014}).

The parameters of the mechanical model is unknown and according to \cite{RBN2007}, parameter values are usually determined in a state where tissues are submitted to gravity forces (the loaded configuration). So, we infer the reference state of the breast (the unloaded configuration) by observing its deformed (loaded) state. In order to determine the material parameters of soft tissues for simulations and to provide the reference configuration, optimisation procedures are usually carried out on the basis of MR imaging or radiological acquisitions such as mammographic plate compression \cite{Az02}, \cite{LSB2003},\cite{PCH2008},\cite{RBN2007},\cite{AFB2015}, \cite{EVH2016}. Breast models have also been assessed using predicted location of anatomical landmarks and selected in breast images acquired before and after {\it in vivo} compression by visual comparison \cite{IIP2016,ERJ2014}. 

The use of nonlinear models to simulate the skin, the muscles, and the tissues, is well-developed in the bio-mechanics context \cite{Gas14},\cite{fung12},\cite{raja04}, \cite{Katartzis2002} and \cite{Esslinger2019} while the finite element method is a popular technique for the discretisation \cite{Py12},\cite{Z05},\cite{Lee12} using alternatively the weak formulation or the minimisation framework \cite{ball83},\cite{Cy98},\cite{Pinci2003}. The Neo-hookean is considered as a well-adapted mechanical representation of the breast \cite{PCH2008}, \cite{Au13} and more generally for human tissues \cite{AFB2015},\cite{IIP2016}.

The methods proposed in literature are based on very complex inverse elasticity problems and image segmentation processes leading to a high computational cost (\cite{HHT2012},\cite{EVH2016}), that would be unnecessary or too expensive for routinely or low-cost practices. From a practical point of view, the objective of the proposed method we develop hereafter, is to recreate a digital model of the breast that correctly represents the behaviour to the real breast (\cite{PCH2008},\cite{RBN2007}) but with simple parameters' identification. \cite{ERJ2014} demonstrates the possibility to use less complex models that provide satisfactory results. and a simpler method like the one presented in \cite{Au13}, can be beneficial in terms of the simulation's performance. 

We propose a new and simple method to estimate biomechanical parameters of a breast model to provide an effective way of retrieving the breast parameters to be, in turn, used to accurately predict breast deformations. This model considers both the geometrical and mechanical parameters including the skin effect leading to a more complete six parameters model (method proposed by \cite{Au13} only considers a two-parameter problem). For the particular case of breast, we differentiate the skin from the core assuming a couple of coefficients for each material. Moreover, an additional goal of the present model is to develop a numerical method that runs on a mid-range laptop and produce results within a few minutes (see also \cite{Lopes2017}). The key ideas are, on the one hand, the introduction of a simple set of {\it in vivo} breast measures easy to achieve (avoiding the use of medical imaging technology), and on the other hand, the introduction of a four-parameters non-linear mechanical model defining on a simplified two-parameters reference configuration (breast in a free stress state). Anthropometric measurements provide the data we use to fit the model parameters (both mechanical and geometrical). 

The model has some specific properties that do not exist in commercial software. For example, the mesh is dynamically adjust following the geometrical parameters we aim at fitting with the measurement. Hence the mesh is itself an unknown of the problem while most of the software use a given mesh to perform the simulation. 

The usage of relevant medical data (measurements obtained during consultations) allied to the simplicity of the biomechanical model, make possible the use of this model effectively in various contexts where medical imaging resources are not available. 

The paper is organised as follows. We present the breast model with its mathematical formulation in Section \ref{sec:breastModelling}, followed by the parameter identification method in Section~3. In Section~4 we show the discretisation of this model and parameter evaluation method while results obtained using synthetic measurements are presented in Section~5. Finally, in Section~6 conclusions are presented.




\section{Breast Modelling}
\label{sec:breastModelling}
The breast is a complex structure constituted of a mass of glandular tissue encased in variable quantities of fat, that account for its characteristic round shape, connected to the skin trough a series of ligaments. All these types of tissues possess different mechanical properties while the proportion of these materials varies with factors like genetics and age \cite{poplack2004}. Moreover, the breast skin plays a major role as a wrapper tissue enjoying specific mechanical properties. 

\begin{figure}[ht]
\includegraphics[width=0.5\textwidth]{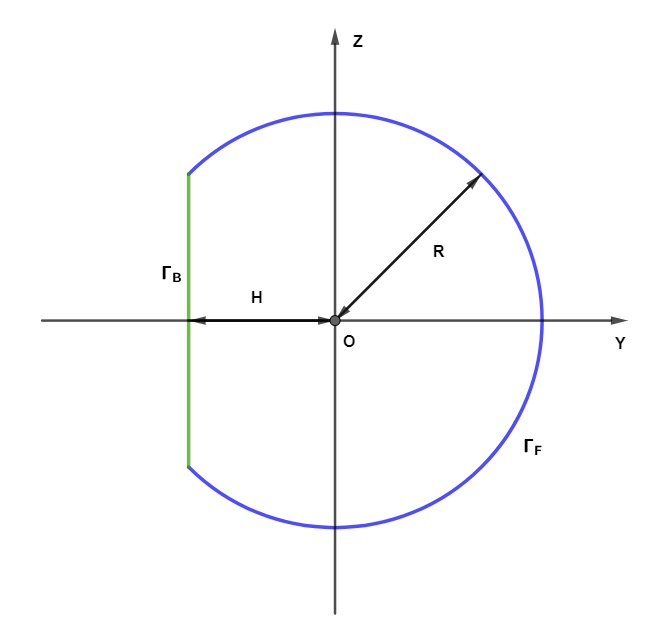}
\centering
\caption{Breast geometry configurations. Radius $R$ characterised the main sphere while $H$ identifies the truncated part following the negative $y$ direction.}
\label{fig:breast2_1}
\end{figure}

Deformation evaluation of the breast over external actions is achieved by considering a simplified stress-free geometrical domain of the breast equipped with the Neo-hookean mechanical model \cite{PCH2008} where we differentiate the breast inner tissues (fat and glandular) and skin. Additionally, we also introduce the Chassaignac space to reproduce the breast mobility \cite{Chass10}.
The gravity-free reference breast domain ${\Omega_{g}}$ is a piece of spherical cap where the plane section is attached to the torso. The domain is parameterised with the radius $R$ of the sphere while $H<R$ represents the non-truncated length as displayed in Figure \ref{fig:breast2_1}. The cap which is placed in the torso plane corresponds to a circle of radius $r$ which is smaller than $R$. We could use a more complex shape such as a super-ellipsoid as in \cite{Kutra2015}, but this shape was proven to be representative as it was shown in \cite{Au13,Lopes2017}.

\begin{figure}[ht]
\includegraphics[width=0.55\textwidth]{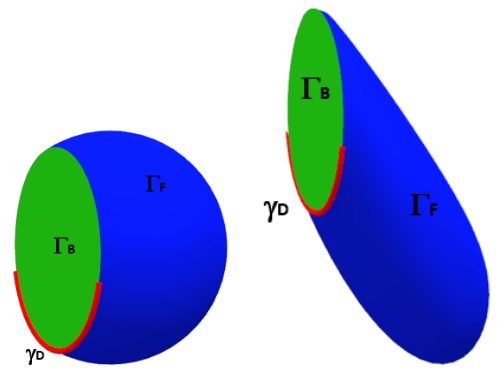}
\centering
\caption{Notation and geometry of the gravity-free geometry (left) and configuration in the gravity field (right)}
\label{fig:breast1}
\end{figure}

The final and more realistic shape of the breast is obtained by determining the effect of gravity on the breast tissues as observed in figure \ref{fig:breast1}.

To prescribe the boundary condition and compute the energy associated to the skin, we introduce the following notations, reproduced on Figure \ref{fig:breast1}: $\Gamma_\text{B}$ is the back side plane of the breast which attaches to the torso, $\Gamma_\text{F}$ represents the surface associated to the skin of the breast and $\gamma_\text{D}$ is the arc of the infra-mammary fold. Like in \cite{Au13,Lopes2017} we use the $\gamma_\text{D}$ as a fixation boundary in order to allow breast mobility.

The hyperelastic neo-Hookean compressible relations \cite{Chass10} equipped with the so-called Chassaignac space represent a well-accepted model for biomechanical soft tissues \cite{Au13}. The Chassaignac space corresponds to a mobile zone $\Gamma_\text{B}$ located between the breast and the trunk which acts as a spring to maintain the breast close to the trunk. Under the gravity, the breast corresponds to a minimisation of the energy functional \cite{ball83,Cy98,Pinci2003} associated with the neo-Hookean system. Such a functional aggregates the internal energy due to the bulk, the energy deriving from the skin displacement, the gravitational energy characterised by the gravitational field $a_{g}\in\mathbb R^3$ and the spring energy associated to Chassaignac space. Anisotropic model for skin \cite{Groves2013} is also investigated to provide a more sophisticated description taking into account the mechanical behaviour of skin under large deformations.

For any generic point $p\in \Omega_g$, we seek the new position vector field $f:p\in \Omega_g\to f(p)\in \mathbb R^3$ which minimises the energy functional given by
\begin{align}
\label{intro-problema3}
{\cal J}(f) &=\int_{{\Omega_{g}}} W_{\text{br}}(\nabla f(p)) \text{d}p+\int_{\Gamma_\text{F}} W_{\text{sk}}(\nabla_{||} f(p))\text{d}S_p- \nonumber\\
& \qquad \int_{\Omega_{g}}\rho \: a_\text{g}\cdot f(p)\text{d}p+\int_{\Gamma_\text{B}} c \|f(p)-p\|\text{d}S_p
\end{align}
subject to the constraints
\begin{align}
& f(p)=p,           && \text{in } \gamma_\text{D},\label{gra:cond1}\\
& f(p).\vec{N}^{B}=0,        && \text{in } \Gamma_\text{B}\label{gra:cond2}
\end{align}
where $W_{br}$ and $W_{sk}$ represents the volume and skin surface strain-energy respectively. $\rho$ stands for the bulk density, $c$ represents the Chassaignac coefficient and $a_{g}$ is represents gravity. Constraint~(\ref{gra:cond1}) represents the fixation of the infra-mammary fold on the torso, i.e., any point $p$ in $\gamma_\text{D}$ maintain their position while constraint~(\ref{gra:cond2}) states that any point $p$ on the breast plane attached to the torso (Chassaignac space) can move laterally but not along the normal direction, \textit{i.e.} $p$ can only move inside that trunk plane.

The energy functional given by equation \ref{intro-problema3} states that each point $p$ affected by $f(p)$ will move according to the internal energy of the bulk ($W_{br}(f(p))$), the skin displacement ($W_{sk}(f(p))$) as well as the gravitational energy ($\rho\ a_{g}\cdot f(p)$) and the energy associated to the Chassaignac Space ($c||f(p)-p||$). Note that, despite modelling the breast bulk tissue and skin differently, the skin is in fact welded to the breast, so each $p \in \Gamma_{F}$ belongs to a inner breast cell.Just as in \cite{PCH2008}, we consider the bulk of the breast tissue as an homogeneous tissue whose viscoelastic properties values are proportional to the amount of fat and glandular tissue.

The expressions for the volume strain-energy and the skin strain-energy densities, respectively represented by $W_{\text{br}}$ and $W_{\text{sk}}$, are given by
\[
W_{\text{br}}(F)=\frac{\mu_{\text{br}}}{2}\left( (FF^t) -3 -2\ln\left(\det(F)\right) \right)+\frac{\lambda_{\text{br}}}{2}\left( \det(F)-1 \right)^2,
\]
where $F=\nabla f$ is the Jacobian matrix of $f$ and $(\lambda_{\text{br}},\mu_{\text{br}})$ are the Lam\'{e} parameters for the breast, and
\[
W_{\text{sk}}(F_{||})=\frac{\mu_{\text{sk}}}{2}\left( (F_{||}F_{||}^t) -2 -2\ln\left(\det(F_{||})\right) \right)+\frac{\lambda_{\text{sk}}}{2}\left( \det(F_{||})-1 \right)^2,
\]
where $(\lambda_{\text{sk}},\mu_{\text{sk}})$ are the Lam\'{e} parameters for the skin while $F_{||} = \nabla_{||} f_{||}$ is the 
Jacobi matrix of the superficial (skin) displacement $f_{||}$ ($f_{||}$ is the restriction of $f$ on $\Gamma_\text{F}$ using local two-parameters
representation since $\Gamma_\text{F}$ is a surface).

\section{Parameters identification}
\label{sec:3}
Let denote $\Lambda=(R,H,\lambda_{br}, \mu_{br},\lambda_{sk},\mu_{sk})$ the six parameters which represents the geometrical $\Lambda_{g}=(R,H)$ and the physical degrees of freedom $\Lambda_{m}=(\lambda_{br}, \mu_{br},\lambda_{sk},\mu_{sk})$ respectively. First, for the two geometrical parameters, we create the stress-free configuration $\Omega_{g}$, as shown in Figure \ref{fig:breast2_1}, which defines the operator
\begin{align}
\label{parameters-1}
\Lambda_{g} \to \Omega_{g}=\Omega_{g}(\Lambda_{g}).
\end{align}
Secondly, for the four physical parameters, minimisation of the energy $\mathcal J(f)=\mathcal J(f;\Omega_{g},\Lambda_{m})=\mathcal J(f;\Lambda)$ over the set of regular functions $f$ defined on $\Omega_{g}$ provides the solution $f(p)=f(p;\Lambda)$. At last, applying the displacement over the reference domain provides the deformed domain $\Omega(\Lambda)=f(\Omega_g)$ which, at the end of the day, defines the operator
\begin{align}
\label{parameters-2}
\Lambda\to (\Lambda_{m},\Omega_{g}) \to \Omega(\Lambda)=f(\Omega_g) = \big \{f(p,\Lambda);\ p\in \Omega_g \big \}.
\end{align}

\begin{figure}[ht]
\includegraphics[width=0.60\textwidth]{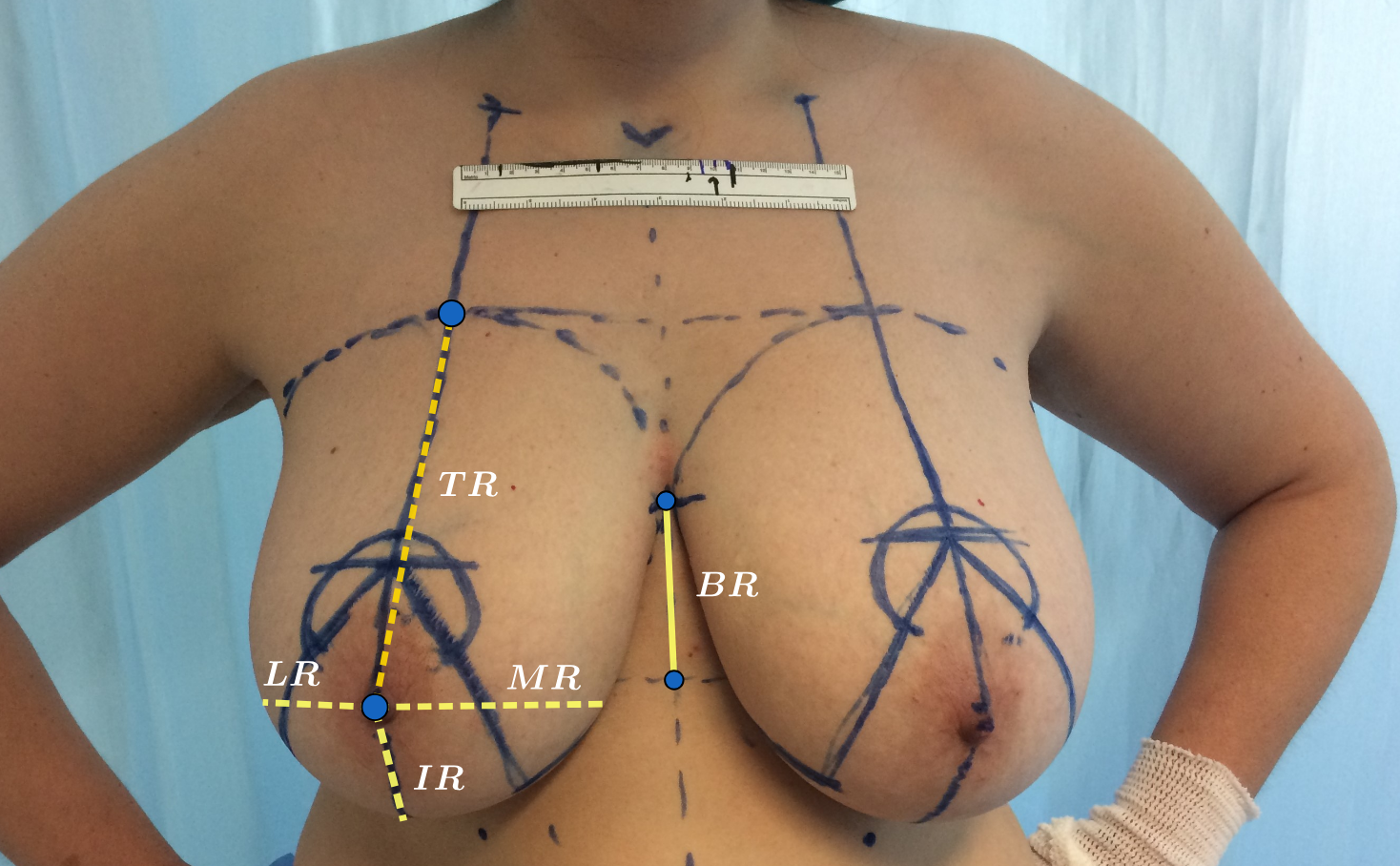}
\centering
\caption{Markings on a patient to plan a breast reduction surgery. Photography taken by Dr. Augusta Cardoso and reproduced with her kind permission.}
\label{fig:breast_the_marks}
\end{figure}

\begin{figure}[ht]
\includegraphics[width=0.31\textwidth]{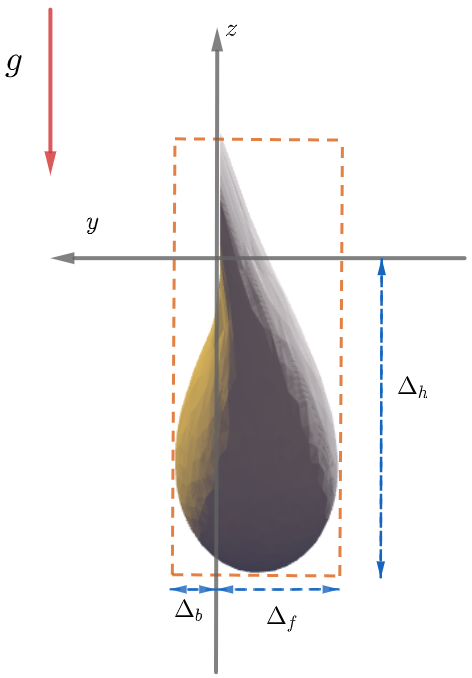}
\includegraphics[width=0.38\textwidth]{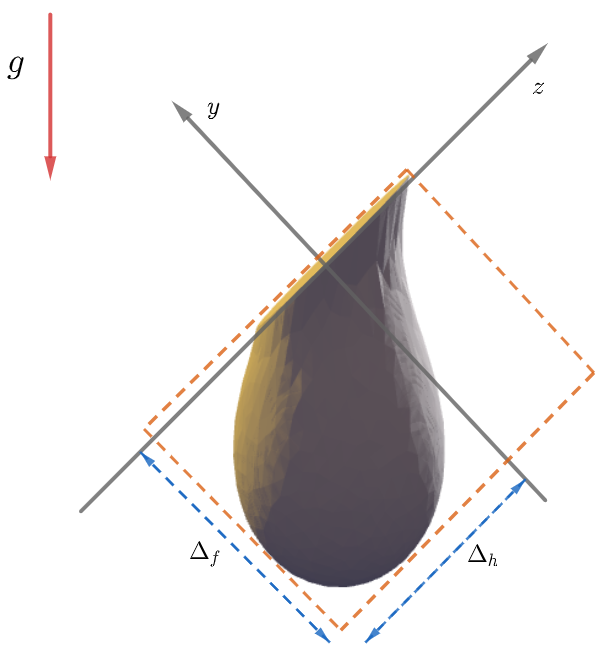}
\includegraphics[width=0.31\textwidth]{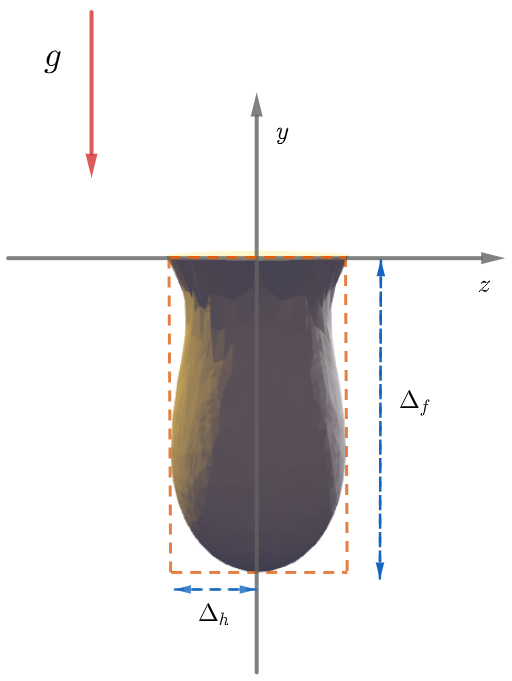}
\includegraphics[width=0.38\textwidth]{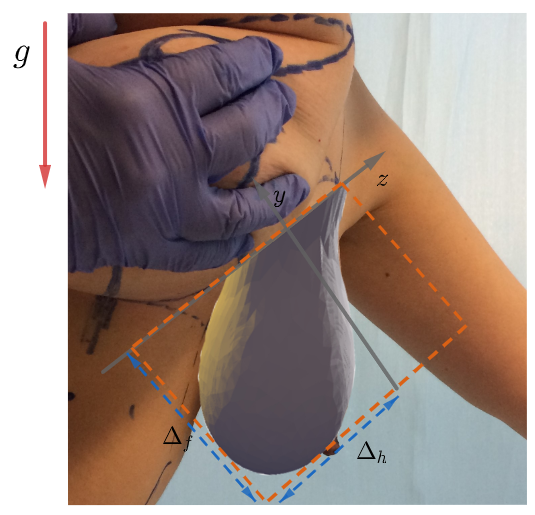}
\centering
\caption{Determination of the breast height, frontal and back depth for a patient in different positions (top left - standing up; top right - bend; bottom left - lying down; bottom right - real example of a patient bending forward). We determine the bounding boxes with an axis aligned to the patient's torso. Photography taken by Dr. Augusta Cardoso and reproduced with her kind permission.}
\label{fig:breast_eval1}
\end{figure}

Third, to perform the parameters identification, we consider a set of measurements based on the the markings performed by surgeons when planning breast surgeries (figure \ref{fig:breast_the_marks}) as well as others mentioned on literature. The goal is to use anthropometric measurements that a surgeon can perform during a consultation before the surgery and then use them to retrieve the breast biomechanical parameters. 
We consider the following measurements: 
\begin{itemize}
    \item Breast Volume ($volume$): calculated with the breast inferior radius $IR$, the lateral radius $LR$, the medial radius $MR$ and the breast depth is given by $min_{y}$ using a formula presented in \cite{Kayar2011};
    \item Skin surface area ($area$): calculated with the same elements as the volume but adding the breast base radius $BR$ and the distance between the top of the breast and the nipple ($TR$);
    \item Height of the breast ($\Delta_{h}$): difference between the highest point and the lowest point of the breast (see figure \ref{fig:breast_eval1}); 
    \item Frontal Depth ($\Delta_{f}$): distance between the torso and the most frontal point of the breast (can be obtained with a normal or squared ruler - figure \ref{fig:breast_eval1});
    \item Back Depth ($\Delta_{b}$): distance between the torso and the back of the breast (see figure \ref{fig:breast_eval1}).
\end{itemize}
While the breast volume and skin surface area are obtained using formulas, the other three measures are obtained by determining a bounding box that encases the breast as figure \ref{fig:breast_eval1} shows (top left). 

We have 5 measures but 6 parameters hence to provide a consistent non-singular inverse problem, we carry out the  measurements with the patient in three different positions (standing up, bending forward approximately 45º and on all fours) that provide 15 measures denoted $\overline{M}=(\overline{M}_{1},\cdots,\overline{M}_{15})^T$. Notice that the bounding boxes that characterise the referential for the three positions are aligned along the breast base (axis $Oz$ and $Ox$), \textit{i.e.}, changes with the patient's position. The measures are performed with respect to the  bounded boxed as indicated in figure \ref{fig:breast_eval1}.

For a given a set of parameters, we solve the direct problem corresponding to the three positions. From the deformed configuration, we then compute the vector $M(\Lambda)=M(\Omega(\Lambda))=(Mk_{1},\cdots,M_{15})^T$. Conversely, the inverse elasticity problem consists in seeking the set of parameters such that $M(\Lambda)$ is equal to $\overline{M}$ in the least-square sense.
Introducing the cost function
\begin{align}
\label{optimise-problema1}
E(\Lambda) = \sum_{i=1}^{n}(M_{i}(\Lambda) - \overline{M_{i}})^{2}
\end{align}
we seek for $\overline{\Lambda}=\arg \min_{\Lambda} E(\Lambda)$ which minimises the errors between 
the experimental data and the theoretical solution values. Because the measurements we use are different (the volume is measured in cubic meters, the area is measured in squared meters and the other three are measured in meters) we consider relative values for these measurements in order to give the same importance to every measure. Hence we rewrite equation \ref{optimise-problema1} as
\begin{align}
\label{optimise-problema1_mi}
E(\Lambda) = \sum_{i=1}^{n}(m_{i}(\Lambda) - \overline{1_{i}})^{2},
\end{align}
where $m_{i}(\Lambda)= \frac{M_{i}(\Lambda)}{\overline{M_{i}}}$.
To provide the optimal set of parameters, an iterative process is considered by evaluating the sequences $\Lambda^{k+1}=\Lambda^{k}+\Delta\Lambda^{k}$ such that $E(\Lambda^{k+1})<E(\Lambda^{k})$. We use the Gauss-Newton method to take advantage that function $E$ casts in the specific form
\begin{align}
\label{optimise-problema2}
E(\Lambda) = (m_{i}(\Lambda) - \overline{1_{i}})^{T}(m_{i}(\Lambda) - \overline{1_{i}}).
\end{align}
The first-order Taylor series expansion reads 
\begin{align}
\label{optimise-problema3}
m(\Lambda+\Delta\Lambda) \approx m(\Lambda) + \frac{\partial m}{\partial\Lambda}(\Lambda)\, \Delta\Lambda = 
m +  \frac{\partial m}{\partial\Lambda}\, \Delta\Lambda
\end{align}
where $\displaystyle \frac{\partial m}{\partial\Lambda}$ is the Jacobian matrix. From relations \ref{optimise-problema2} and \ref{optimise-problema3}
we deduce the first term of error for a perturbation of the parameters $\Delta E=E(\Lambda + \Delta\Lambda)-E(\Lambda)$
\begin{align*}
\begin{split}
\Delta E= - \Delta\Lambda^{T}\Big[2\left (\frac{\partial m}{\partial\Lambda}\right )^{T}
(\overline{1}-m(\Lambda)) -\left (\frac{\partial m}{\partial\Lambda}\right)^{T}\frac{\partial m}{\partial\Lambda}\,\Delta\Lambda \Big].
\end{split}
\end{align*}
Thus, by identification, we deduce
\begin{align}
\label{optimise-problema5}
\frac{\partial E}{\partial\Lambda}(\Lambda)\approx -2\left (\frac{\partial m}{\partial\Lambda}\right)^{T}(\overline{1}-m(\Lambda)) + 
\left(\frac{\partial m}{\partial\Lambda}\right)^{T}\frac{\partial m}{\partial\Lambda}\,\Delta\Lambda
\end{align}
We seek a perturbation $\Delta\Lambda$ that minimises $E$, i.e., $\frac{\partial E}{\partial\Lambda} = 0$. Therefore Equation \ref{optimise-problema5} provides the expression 
\begin{align}
\label{optimise-problema6}
\Delta\Lambda = 2\Big[ \left(\frac{\partial m}{\partial\Lambda}\right)^{T}\frac{\partial m}{\partial\Lambda}\, \Big]^{-1}
\left(\frac{\partial m}{\partial\Lambda}\right)^{T}(\overline{1}-m(\Lambda)) 
\end{align}
Since a direct computation of the jacobian matrix is not possible, we numerically evaluate each partial derivative setting
$$
\partial_{i}M(\Lambda) \approx \frac{M(\Lambda+\varepsilon e_i)-M(\Lambda)}{\overline{M} \varepsilon} \in \mathbb R^n
$$
where $\varepsilon$ is fixed by the user in function of the  minimisation problem and 
$\displaystyle e_i=(\delta_{ij})_{j=1,\cdots, n}$.

\section{Discretisation}
To carry out numerical simulations, we introduce the discretisation of the breast model and the optimisation problem. We denote by $\mathcal T_{h,g}$ a mesh of the gravity free domain $\Omega_{g}=\Omega_{g}(\Lambda_g)$ constituted of $I$ non-overlapping tetrahedron cells $\tau_i$, $i=1,\ldots,I$, and $N$ vertices $P_n=(P_{nx},P_{ny},P_{nz})\in\mathbb R^3$, $n=1,\ldots,N$ while
$$
\Omega_{g,h}=\bigcup_{\tau_i \in \mathcal T_{h,g}}\tau_i
$$
stands for the discrete domain. Moreover, $T_k$, $k=1,\ldots,K$, represents the faces of the tetrahedrons of the mesh that belong to $\Gamma_\text{F}$. Quantities $|\tau_i|$ and $|T_k|$ represent the volume and the area of the cell and the triangle respectively.
We also use a local indexation and denote by  $P_{ij}=(P_{ijx},P_{ijy},P_{ijz})\in\mathbb R^3$, $j=1,2,3,4$, the vertices of $\tau_i$ and by $P_{kj}=(P_{kjx},P_{kjy},P_{kjz})\in\mathbb R^3$, $j=1,2,3$, the vertices of $T_k$.
To discretise function $f$, we associate to each node $P_n$ an approximation $f_n\approx f(P_{n})$ and denote by $f_{h}$ the continuous, linear piecewise function while vector $\Phi_{h}=(f_{nx},f_{ny},f_{nz})_{p}$ collects the $3N$ components of the new positions. 

Vector $\Phi_{h}$ corresponds to the new configuration but not all the entries are necessarily unknowns of the problem since some of them are characterised by the boundary conditions. The sub-vector $X_{h}$ of $\Phi_{h}$ only contains the unknown values we shall use in the minimisation process while the boundary conditions define an operator
\[
X_{h}\to \Phi_{h}=\mathcal B(X_{h})
\]
which provides the other entries to complete vector $\Phi_{h}$.
In the present contribution, condition~\eqref{gra:cond1} yields that $f_n=P_n$ for any $P_n\in \gamma_\text{D}$ while relation \eqref{gra:cond2} implies that for any $P_n\in\Gamma_\text{B}$, we set $f_{ny}=0$ to maintain interface $\Gamma_\text{B}$ on the trunk plane $y=0$.
This two conditions completely define the vector of unknowns $X_{h}$ and operator $\mathcal B$.

\subsection{The energy functional}
Discrete version of the energy functional is given by
$$
J_h(X_h)=J_h(\text{B}(X_h))=\widehat J_h(\Phi_h)=J_h^1+J_h^2+J_h^3+J_h^4
$$
where $J_{h}^{1}$, $J_{h}^{2}$, $J_{h}^{3}$ and $J_{h}^{4}$ represent the volume energy, the surface energy, the energy from the gravitational displacement, and the energy associated to the Chassaignac space respectively. 

$\bullet$ For a new configuration characterised by the approximation $p\in{\Omega_{g}} \to f_h(p)\in\mathbb R^3$ and stored in vector $\Phi_h$, the internal energy on tetrahedron $\tau_i$ is given by
\[
W_{\tau_i}=\vert \tau_i\vert
\left(
\frac{\mu}{2} \Big [\textrm{tr}(F_{i}F_{i}^T)-3-2\ln(\det(F_{i}))\Big]+\frac{\lambda}{2} \Big[\det(F_{i})-1\Big]^2
\right),
\]
where $F_i$ is the $3\times 3$ matrix solution of the linear system
$$
f_{i2}-f_{i1}=F_i(P_{i2}-P_{i1}), \quad f_{i3}-f_{i1}=F_i(P_{i3}-P_{i1}),
$$
$$
\quad f_{i4}-f_{i1}=F_i(P_{i4}-P_{i1}).
$$
finally having 
$\displaystyle J^1_h=\sum_{\tau_i \in \mathcal T_{h,g}} W_{\tau_i}$.

$\bullet$ For the surface energy, the discrete piecewise linear function $f_h$ transforms a triangle $T_k$ with vertices $OAB$ into a triangle $T_k'$ with vertices $O'A'B'$. The method is similar to the surface variation using in \cite{Lee2018} to evaluate the energy deriving from the displacement variations. Since the translation and the rotation do not change the stress due to the deformation, we assume that $O'B'$ is collinear to $OB$ and $A'$ belongs to the same plane as triangle $OAB$. Function $f_{||}$ is a two-dimensional function locally given by $f_{||}(O)=O$, $f_{||}(A)=A'$, $f_{||}(B)=B'$.
The Jacobian matrix of $f_{||}$ is the constant matrix
$$
Jf_{||}=\mathcal A=\left [\begin{array}{cc}
a & b\\
c & d
\end{array} \right ].
$$
To determine the matrix, one writes
$$
\left [\begin{array}{cc}
a & b\\
c & d
\end{array} \right ]
\left [\begin{array}{c}
\Vert OB\Vert  \\
0
\end{array} \right ]=
\left [\begin{array}{c}
\Vert O'B'\Vert  \\
0
\end{array} \right ], 
$$
$$
\quad
\left [\begin{array}{cc}
a & b\\
c & d
\end{array} \right ]
\left [\begin{array}{c}
\Vert OA\Vert \cos(\alpha)  \\
\Vert OA\Vert \sin(\alpha)
\end{array} \right ]=
\left [\begin{array}{c}
\Vert O'A'\Vert \cos(\alpha')  \\
\Vert O'A'\Vert \sin(\alpha')
\end{array} \right ],
$$
where $\alpha=\angle(OA,OB)$ and $\alpha'=\angle(O'A',O'B')$.
The first linear system gives $c=0$ and $a=\frac{\Vert O'B'\Vert}{\Vert OB\Vert}$. Substituting these 
expressions in the second linear system we obtain
$$
d=\frac{\Vert O'A'\Vert \sin(\alpha')}{\Vert OA\Vert \sin(\alpha)}, \quad
b=\frac{\Vert O'A'\Vert \cos(\alpha')-\frac{\Vert O'A'\Vert}{\Vert OA\Vert}\Vert OA\Vert \cos(\alpha)}{\Vert OA\Vert \sin(\alpha)}.
$$

The superficial energy on triangle $T$ for the skin ($W_{T}$) is then given by
$$
\vert T\vert \left (
\frac{\mu_\text{sk}}{2} \Big [\textrm{tr}(Jf_{||}Jf_{||}^T)-2-2\ln(\det(Jf_{||}))\Big]+
\frac{\lambda_\text{sk}}{2} \Big[\det(Jf_{||})-1\Big]^2
\right )
$$
and the whole superficial energy is approximated by
$$
J_h^2=\sum_{T\in \Gamma_F} W_{T}.
$$

$\bullet$ For $J_{h}^{3}$ and $J_{h}^{4}$ we have 
$$
J_h^3=-\sum_{\tau_i} \frac{|\tau_i|}{4}\rho a_\text{g} \left ( f(P_{i1})+f(P_{i2})+f(P_{i3})+f(P_{i4})\right )
$$
and
$$
J_h^4=\sum_{T_j \subset \Gamma_B} \frac{|T_j|}{3}c \left (\Vert f(P_{i1})-P_{i1} \Vert
+\Vert f(P_{i2})-P_{i2}\Vert +\Vert f(P_{i3})-P_{i3}\Vert \right ). 
$$

\subsection{Numerical approximation of the discrete energy minimiser}\label{sec::minimisation}
For a given vector $X_h$ and the boundary conditions, we deduce vector $\Phi_h=\mathcal B(X_h)$, hence
the continuous linear piecewise function $f_h$ we use to compute the discrete energy functional.
We then build the operator
$$
X_h\to \Phi_h=\mathcal B(X_h) \to J_h(X_h;\Lambda)=\widehat J_h(\Phi_h;\Lambda)\in \mathbb R
$$
where $\Lambda$ is a given set of parameters. The numerical solution we seek provides the vector $\Phi_h=\Phi_h(\Lambda)$
which minimises the energy of the discrete mechanical system. 
Conjugate gradients method is employed to determine the minimiser $\bar X_h$ of the discrete functional $J_h(X_h;\Lambda)$. 

\subsection{Cost function discretisation}\label{sec::optimisation_minimisation}
Let $\Lambda=(\Lambda_{m},\Lambda_{g})$ be a set of parameters. We deduce the discrete gravity free configuration $\Omega_{g,h}$ which provides the operator $\Lambda_{g} \to \Omega_{g,h}$. Then we compute the solution $\Phi_h(\Lambda)$ minimising the discrete energy functional $\mathcal J_{h}(f_{h}) = \mathcal J_{h}(f_{h};\Omega_{g,h},\Lambda_{m}) = \mathcal J_{h}(f_{h};\Lambda)$.
We deduce the final discrete breast $\Omega_{h}(\Lambda)$ after applying the deformation
$$
\Omega_{h}(\Lambda) = \{f_{h}(p,\Lambda); p\in \Omega_{g,h}\}
$$
with $f_{h}$  being the function that provides the new position. We assess the measures on $\Omega_{h}(\Lambda)$ and produce vector $M(\Lambda;h)$. In conclusion, we have define the discrete measures operator $\Lambda \to M(\Lambda;h)$. On the other hand, the discrete cost function reads
\begin{equation}
E(\Lambda;h) = \sum_{i=1}^{n}\Big (M_{i}(\Lambda;h)-\overline{M_{i}}\Big )^{2} \label{eq:cost_function}
\end{equation}
We apply the iterative process to the discrete versions to get the best parameter set $\Lambda$ by calculating successive variation of the parameters ($\Delta \Lambda^n$) until $E(\Lambda^{n};h)-E(\Lambda^{n+1};h) < \epsilon_{m}$ for a given threshold $\epsilon_{m}$.


\section{Synthetic Cases}

\begin{figure}[htb]
\includegraphics[height=6.0cm]{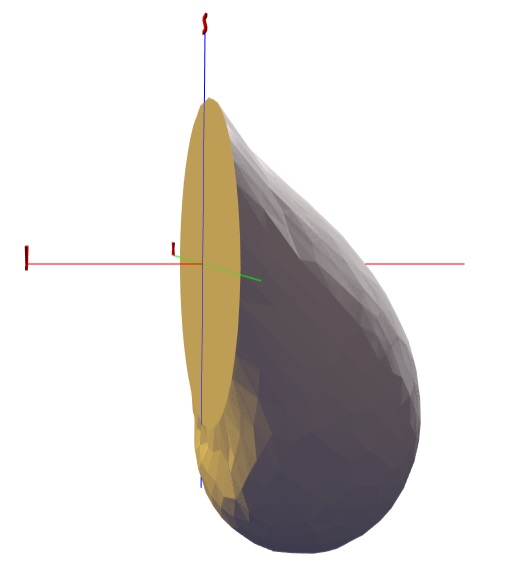}
\centering
\caption{Breast under gravity with the given set of defined parameters $\overline{\Lambda}$.}
\label{fig:breast3}
\end{figure}

To assess the model quality and robustness, several numerical experiences are carried out using a manufactured solution. We solve the direct problem by prescribing a given set of parameters and compute the associated measurements and test the method ability to recover the initial parameters independently of the initial guess. To this end, we create a numerical breast setting the $6$ parameters 
$\overline{\Lambda}=(R = 0.0562m,H = 0.05m,\lambda=1000Pa,\mu=150Pa,\lambda_p=8000Pa,mu_p=1600Pa)^T$  and compute the three configurations in the gravitational fields (see Figure \ref{fig:breast3} for the case stand up) that provides the reference set of measurements 
$$
M(\overline{\Lambda};h)=\overline M_{\bar h}=(\overline M_{\bar h,1},\cdots,\overline M_{\bar h,15})^T
$$ 
depending on the mesh characteristic size $\bar h$ that corresponds to the mesh with 2948 vertices and about 12000 tetrahedrons. For the sake of simplicity, we use the notation $M(\Lambda)$ without mentioning the dependency of the mesh size.
We want to evaluate the importance of the initial guess in the final results with initial guesses that went from 10\% (approximate guess), 30\% (reasonable guess), 60\% (bad guess) and over $60\%$ from the reference parameters. 
The results (mean values) can be seen in table \ref{table:converse_test1}.
\begin{table}[ht]
\centering
\footnotesize
\begin{tabular}{|c|c|c|c|c|c|c|}
\hline
\multicolumn{7}{|c|}{Parameter Loss}                                                                                                                                     \\ \hline
\multirow{2}{*}{\begin{tabular}[c]{@{}c@{}}Mesh Size\\ \& Parameters\end{tabular}} & \multicolumn{2}{c|}{Geometrical} & \multicolumn{4}{c|}{Mechanical}                \\ \cline{2-7} 
                                                                                     & R               & H              & $\lambda$ & $\mu$  & $\lambda_{p}$ & $\mu_{p}$ \\ \hline
\begin{tabular}[c]{@{}c@{}}Coarse \\ (N=226)\end{tabular}                            & 0,08\%          & 0,39\%         & 0,42\%    & 0,89\% & 0,27\%        & 2,77\%    \\ \hline
\begin{tabular}[c]{@{}c@{}}Medium \\ (N=572)\end{tabular}                            & 0,10\%          & 0,47\%         & 1,39\%    & 0,79\% & 0,36\%        & 2,40\%    \\ \hline
\begin{tabular}[c]{@{}c@{}}Medium-Thin \\ (N=1014)\end{tabular}                      & 0,05\%          & 1,20\%         & 1,12\%    & 0,79\% & 1,16\%        & 2,32\%    \\ \hline
\begin{tabular}[c]{@{}c@{}}Thin \\ (N=1404)\end{tabular}                             & 0,06\%          & 0,98\%         & 1,07\%    & 0,76\% & 0,95\%        & 2,28\%    \\ \hline
\begin{tabular}[c]{@{}c@{}}Very-Thin \\ (N=2127)\end{tabular}                        & 0,08\%          & 0,68\%         & 0,92\%    & 0,65\% & 0,85\%        & 2,14\%    \\ \hline
\begin{tabular}[c]{@{}c@{}}Ultra-Thin\\ (N=2498)\end{tabular}                        & 0,07\%          & 0,72\%         & 0,97\%    & 0,62\% & 0,86\%        & 2,13\%    \\ \hline
\end{tabular}
\caption{Values of the parameter loss (in \%) with different sized meshes after 25 iterations. For each mesh size, we selected 10 tests and the values shown are the means from those tests. }
\label{table:converse_test1}
\end{table}
These results show the effectiveness of the algorithm in terms of converging towards the reference parameters with a value of loss close and in some cases inferior to $1\%$ with the exception of the $\mu_{p}$ parameter with a loss value around $2\%$. 

\subsection{Robustness}
\label{subsec:robustness}


We evaluate the computational effort of the method, particularly when increasing the number of vertices. We report the mean squared error (residual) between the reference measurements and the measurements obtained with different mesh sizes with the reference parameters (10 cases per mesh size). We also report the relative running time with respect to the coarser mesh, in function of the mesh size in table \ref{table:granularity_test2}.
\begin{table}[ht]
\centering
\begin{tabular}{|c|c|c|c|c|c|}
\hline
\textbf{Mesh size}    & \textit{\textbf{226}} & \textit{\textbf{572}} & \textit{\textbf{1014}} & \textit{\textbf{1404}} & \textit{\textbf{2127}} \\ \hline
\textbf{Residual}     & 1.52e-3               & 2.72e-4               & 7.26e-5                & 3.91e-5                & 9.13e-7                \\ \hline
\textbf{running time} & 1                     & 3.4                   & 7.8                    & 11.2                   & 20                     \\ \hline
\end{tabular}
\caption{Measurement's difference residual between the reference measurements and the measurements obtained with courser meshes with the same values for the breast parameters $\Lambda$ and the work effort which is spent by the optimisation process}
\label{table:granularity_test2}
\end{table} 
Table \ref{table:granularity_test2} shows that using a thinner mesh with 2127 nodes is 20 times more expensive than a coarser mesh with only 226 elements. Note that the average time of a simulation using a coarse mesh using an intel-i5 2450m processor with 2.3Ghz rounds the 15min.

Notice that the measurements obtained with the reference parameters and a mesh with just 572 nodes are approximately 5\% different when comparing with the reference measurements. This means that by using coarser meshes we are, synthetically, introducing error to the measures. 

\begin{figure}[ht]
\centering
\includegraphics[width=1.0\textwidth]{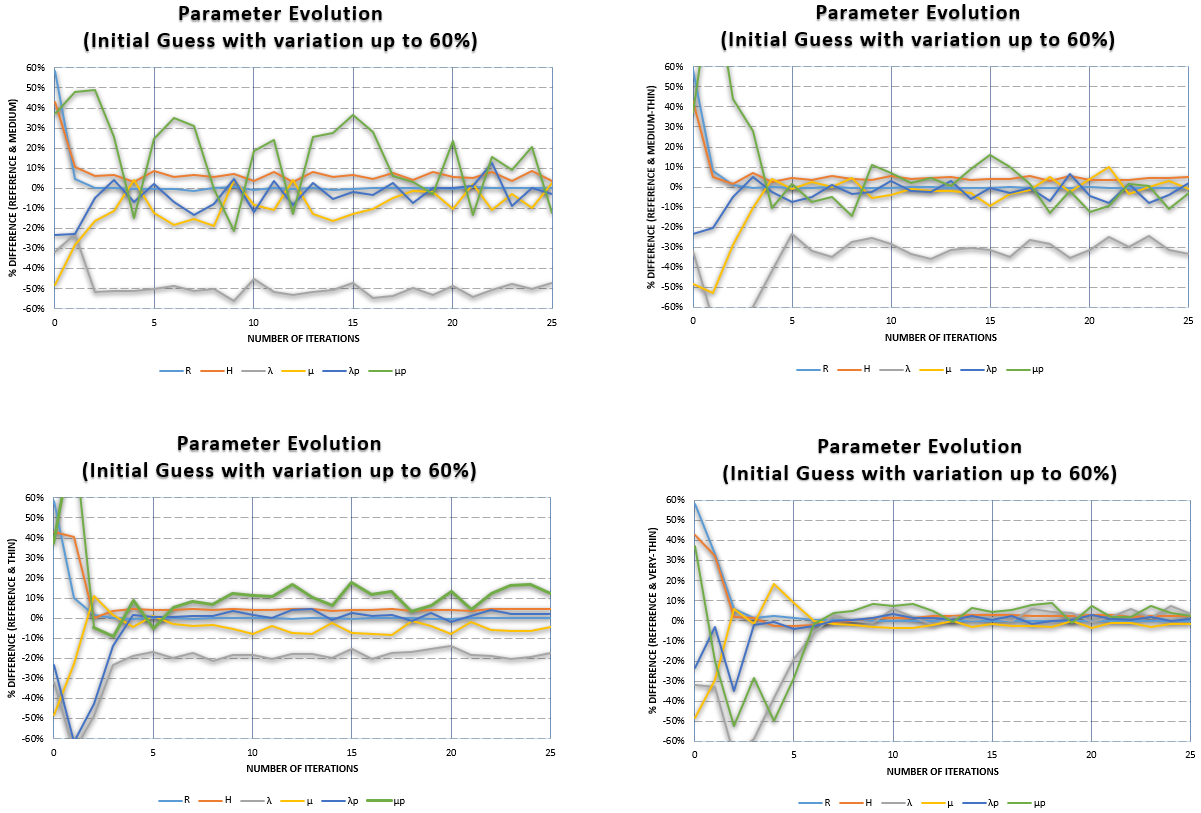}
\caption{Parameter evolution of the optimisation process for different meshes and comparison with the reference values. Top panel corresponds to the coarse mesh of 572 vertices while bottom panel is obtained with 2127 vertices. The intermediate panels correspond to the meshes with 1014 and 1404 vertices.}
\label{fig:graph_parameter}
\end{figure}

We evaluate the precision for different size meshes (namely  medium, medium-thin, thin and very-thin corresponding to a number of 572, 1014, 1404, and 2127 nodes respectively) and we report the values in figure \ref{fig:graph_parameter}.
We observe the volatility for some parameters during the first estimation steps but we obtain stable approximations after iteration 15. Tables \ref{table:granularity_test} and \ref{table:granularity_test_sd} give the mean and standard deviation value of each parameter for those tests in the interval between iteration 15 and 25.
\begin{table}[ht]
\centering
\footnotesize
\begin{tabular}{|c|c|c|c|c|c|c|}
\hline
\multicolumn{7}{|c|}{Parameter Values (Mean)}                                                                                                                            \\ \hline
\multirow{2}{*}{\begin{tabular}[c]{@{}c@{}}Mesh Size\\ \& Parameters\end{tabular}} & \multicolumn{2}{c|}{Geometrical} & \multicolumn{4}{c|}{Mechanical}                \\ \cline{2-7} 
                                                                                     & R               & H              & $\lambda$ & $\mu$  & $\lambda_{p}$ & $\mu_{p}$ \\ \hline
\begin{tabular}[c]{@{}c@{}}Medium \\ (N=572)\end{tabular}                            & 0,35\%          & 6,26\%         & 48,70\%   & 8,71\% & 2,17\%        & 20,33\%   \\ \hline
\begin{tabular}[c]{@{}c@{}}Medium-Thin \\ (N=1014)\end{tabular}                      & 0,24\%          & 4,30\%         & 30,18\%   & 6,82\% & 2,19\%        & 14,74\%   \\ \hline
\begin{tabular}[c]{@{}c@{}}Thin \\ (N=1404)\end{tabular}                             & 0,09\%          & 4,25\%         & 17,93\%   & 5,57\% & 1,64\%        & 14,30\%   \\ \hline
\begin{tabular}[c]{@{}c@{}}Very-Thin \\ (N=2127)\end{tabular}                        & 0,05\%          & 2,53\%         & 4,21\%    & 2,23\% & 1,20\%        & 5,35\%    \\ \hline
\end{tabular}
\caption{Parameter value difference (in \%) between the reference parameters ($\overline{\Lambda}$) and the parameter values obtained with courser meshes after the optimisation process with an initial guess with a variance up to $60\%$ from $\overline{\Lambda}$}
\label{table:granularity_test}
\end{table}
\begin{table}[ht]
\centering
\small
\begin{tabular}{|c|c|c|c|c|c|c|}
\hline
\multicolumn{7}{|c|}{Parameter Values (Standard Deviation)}                                                                                                              \\ \hline
\multirow{2}{*}{\begin{tabular}[c]{@{}c@{}}Mesh Size\\ \& Parameters\end{tabular}} & \multicolumn{2}{c|}{Geometrical} & \multicolumn{4}{c|}{Mechanical}                \\ \cline{2-7} 
                                                                                     & R               & H              & $\lambda$ & $\mu$  & $\lambda_{p}$ & $\mu_{p}$ \\ \hline
\begin{tabular}[c]{@{}c@{}}Medium \\ (N=572)\end{tabular}                            & 0,27\%          & 1,83\%         & 4,55\%    & 7,22\% & 5,01\%        & 21,15\%   \\ \hline
\begin{tabular}[c]{@{}c@{}}Medium-Thin \\ (N=1014)\end{tabular}                      & 0,14\%          & 0,64\%         & 3,66\%    & 4,49\% & 3,53\%        & 12,19\%   \\ \hline
\begin{tabular}[c]{@{}c@{}}Thin \\ (N=1404)\end{tabular}                             & 0,05\%          & 0,31\%         & 2,94\%    & 2,52\% & 1,68\%        & 7,10\%    \\ \hline
\begin{tabular}[c]{@{}c@{}}Very-Thin \\ (N=2127)\end{tabular}                        & 0,05\%          & 0,23\%         & 2,89\%    & 1,53\% & 1,34\%        & 4,01\%    \\ \hline
\end{tabular}
\caption{Parameter value standard deviation (in \%) between the reference parameters and the parameters estimated with courser meshes}
\label{table:granularity_test_sd}
\end{table}

Three levels/rates of convergence are highlighted:
\begin{itemize}
    \item High convergence. The geometrical parameters $\Lambda_{g}=(R,H)$ quickly converge to the solution after few iterations, and present a very stable behaviour (standard deviation values inferior to 2\%). 
    \item Mild convergence. The estimation method presents good approximations for the physical parameters $\mu$ and $\lambda_{p}$.
    \item Rough convergence. Parameters $\lambda$ and $\mu_{p}$ approximation is far away when using coarse meshes and converge to a wrong solution after some few iterations. 
\end{itemize}
The results show that using a very coarse mesh, the algorithm is not capable of determining with accuracy the parameter $\lambda$ and because of that, the convergence of the other parameters is more erratic. That is to be expected because, as mentioned previously, using coarser mesh results in introducing an error in the measurements, \textit{i.e.}, $M(\Lambda)_{coarse}$ and $M(\Lambda)_{thin}$ produce different measurements but $M(\Lambda)_{thin}$ is closer to $\overline{M}$ than $M(\Lambda)_{coarse}$. So, by using more refined meshes, this difference is mitigated and therefore the algorithm is capable of converging more efficiently.

\subsection{Sensitivity Analysis}
Sensitivity analysis is an important issue for assessing the robustness of the method and evaluating the measures that preponderantly influence the parameters. Moreover, potential errors might derive from the performed measurements in a practitioner's office and one has to assess the consequences of such errors on the parameters. At last, it is a good indicator of ill-posed inverse problem. Minimising the cost function $\Lambda \to E(\Lambda;h)$ provides the relation $\overline{M}\to \Lambda(\overline{M};h)$, {\it i.e.} the best parameters that fit the measurements. For the sake of simplicity we drop the reference to h, and so we denote $\Lambda(M)$. The relative  sensitivity analysis aims at assessing the impact of the measure variations on the parameters evaluation. Low relative sensitivity of parameter $R$ with respect to measure $m_1$ means that the relative partial derivative
$$
d_{R,m_1}=\frac{\overline{m_{i}}}{R(\overline{M})}\frac{\partial R}{\partial m_1}(\overline{M})<<1
$$
while an high sensitivity is obtained if $d_{R,m_1}>>1$.

We assume that practitioners provide measurements $m_{i}$ with a relative error of $10\%$ and therefore, to simulate the error impact, we define two new sets of parameter $\Lambda(m_{i}+)$ and $\Lambda(m_{i}-)$ corresponding to a positive and a negative variation of $10\%$ of parameter $\overline{m}_i$, the others being fixed. To calculate the sensitivity value we approximate the relative derivative with the expression $j=1,\cdots,6$ and $i=1,\cdots,15$
\begin{equation}
    \frac{\overline{m_{i}}}{\Lambda_{j}(\overline{M})}\frac{\partial\Lambda_j}{\partial m_i}(\overline{M}) \approx  \frac{1}{\Lambda_{j}(\overline{M})} \frac{\Lambda_j(M_{i}+) - \Lambda_j(M_{i}-)}{0.2}, \qquad 
    M_{i} \pm = \overline{M} \pm 0.1 \left[\begin{array} {c}
                                                0  \\
                                                0 \\
                                                \overline{m_{i}} \\
                                                0 \\
                                                0 
                                                
                                    \end{array}\right]
\end{equation}
which results in a $6\times 15$ matrix that approximate the Jacobi matrix. Computations are achieved with the very-thin mesh and we report the results in Table \ref{fig:sensibility}  with the parameters $\Lambda_j$ in row and the measures $m_i$ in columns. 

\begin{remark}
Such data have to be handled and interpreted with caution since we compute a numerical approximation of the derivative using variations of order $\pm 10\%$. Computational errors due to the discretisation may degrade the approximation accuracy.
\end{remark}


\begin{table}[!ht]
\setlength{\tabcolsep}{5pt}
\renewcommand{\arraystretch}{1.15}
\centering
\begin{tabular}{|c|c|c|c|c|c|c|}
\hline
                  & \cellcolor[HTML]{9AFF99}{\color[HTML]{333333} $\mathbf{R}$} & \cellcolor[HTML]{96FFFB}\textbf{$\mathbf{H}$} & \cellcolor[HTML]{FFCE93}$\boldsymbol{\lambda}$ & \cellcolor[HTML]{CBCEFB}$\boldsymbol{\lambda}\mathbf{_{p}}$ & \cellcolor[HTML]{FFCCC9}$\boldsymbol{\mu}$ & \cellcolor[HTML]{ECF4FF}$\boldsymbol{\mu}\mathbf{_{p}}$ \\ \hline
$\mathbf{m_{1}}$  & 0.0016                                                      & 0.0113                                        & 0.0478                                                       & 0.0058                                                       & 0.0272                                                   & 0.0021                                                   \\ \hline
$\mathbf{m_{2}}$  & 0.2628                                                      & 0.2802                                        & \cellcolor[HTML]{FFCE93}5.4429                               & 1.0576                                                       & \cellcolor[HTML]{FFCCC9}8.2985                           & 1.4528                                                   \\ \hline
$\mathbf{m_{3}}$  & 0.0566                                                      & 0.5668                                        & 1.5790                                                       & \cellcolor[HTML]{CBCEFB}4.3696                               & 1.2637                                                   & \cellcolor[HTML]{ECF4FF}12.3298                          \\ \hline
$\mathbf{m_{4}}$  & 0.3232                                                      & \cellcolor[HTML]{96FFFB}0.8874                & \cellcolor[HTML]{FFCE93}17.4148                              & 3.0257                                                       & \cellcolor[HTML]{FFCCC9}5.4678                           & \cellcolor[HTML]{ECF4FF}10.6005                          \\ \hline
$\mathbf{m_{5}}$  & 0.0369                                                      & 0.2766                                        & 0.6299                                                       & 1.1390                                                       & 0.0739                                                   & 3.4336                                                   \\ \hline
$\mathbf{m_{6}}$  & 0.0016                                                      & 0.0034                                        & 0.0611                                                       & 0.0128                                                       & 0.0157                                                   & 0.0251                                                   \\ \hline
$\mathbf{m_{7}}$  & 0.4995                                                      & 0.7457                                        & \cellcolor[HTML]{FFCE93}7.8666                               & \cellcolor[HTML]{CBCEFB}5.0323                               & \cellcolor[HTML]{FFCCC9}7.0629                           & \cellcolor[HTML]{ECF4FF}18.0483                          \\ \hline
$\mathbf{m_{8}}$  & 0.1920                                                      & 0.1134                                        & \cellcolor[HTML]{FFCE93}16.4224                              & \cellcolor[HTML]{CBCEFB}5.0502                               & 2.6954                                                   & \cellcolor[HTML]{ECF4FF}23.8555                          \\ \hline
$\mathbf{m_{9}}$  & \cellcolor[HTML]{9AFF99}1.0510                              & 0.2342                                        & \cellcolor[HTML]{FFCE93}5.2028                               & 0.1780                                                       & 0.8115                                                   & 1.2001                                                   \\ \hline
$\mathbf{m_{10}}$ & 0.0038                                                      & 0.0317                                        & 0.0846                                                       & 0.0549                                                       & 0.0025                                                   & 0.1724                                                   \\ \hline
$\mathbf{m_{11}}$ & 0.0022                                                      & 0.0063                                        & 0.0359                                                       & 0.0144                                                       & 0.0001                                                   & 0.0416                                                   \\ \hline
$\mathbf{m_{12}}$ & 0.0599                                                      & 0.6822                                        & 0.7975                                                       & 0.5324                                                       & 2.5195                                                   & 0.4907                                                   \\ \hline
$\mathbf{m_{13}}$ & 0.3179                                                      & \cellcolor[HTML]{96FFFB}1.7219                & \cellcolor[HTML]{FFCE93}5.7136                               & 2.4447                                                       & 0.2099                                                   & \cellcolor[HTML]{ECF4FF}7.8664                           \\ \hline
$\mathbf{m_{14}}$ & \cellcolor[HTML]{9AFF99}0.9880                              & 0.1418                                        & \cellcolor[HTML]{FFCE93}4.2689                               & \cellcolor[HTML]{CBCEFB}4.4837                               & 0.6020                                                   & \cellcolor[HTML]{ECF4FF}11.0254                          \\ \hline
$\mathbf{m_{15}}$ & 0.0020                                                      & 0.0121                                        & 0.1050                                                       & 0.0077                                                       & 0.0191                                                   & 0.0355                                                   \\ \hline
\end{tabular}
\caption{Table with the sensibility analysis of the parameters. The coloured cells mark the most relevant measurements for each parameter.}
\label{fig:sensibility}
\end{table}

Table \ref{fig:sensibility} reports the sensitivity coefficients and we highlight with different colours the most relevant measurements for each parameter. All the values are positive since the parameters increase with respect to the measures. We also observe that measurements $m_{1}$, $m_{6}$, $m_{10}$, $m_{11}$ and $m_{15}$ have small impact (even no impact) on the parameter evaluation corresponding to a very low sensitivity. The table also suggests that the choice of measurement to assess the parameters is relevant since the matrix is of maximum rank. 


\section{Conclusion}
We have proposed a new model to simulate breast displacement and identify the mechanical parameters associated to the Neo-Hookean model. The main points are the introduction of a two-parameter geometry for the gravity-free configuration allied to a four-parameter mechanical property model as well as a user friendly set of measurements that does not require sophisticated equipment and turns to be adequate for routinely operations. The sensitivity study with respect to the measure and the mesh has been carried out in a synthetic context to demonstrate the robustness of the method and its capacity to retrieve the good parameters with a controlled range of error. The results show a model capable of estimating with accuracy the parameters of the breast. The simplicity of the model allied to the usage of relevant medical data allows to obtain satisfactory results in a small time period using a mid range laptop.


\section{Acknowledgements}

A big thanks to Dra. Augusta Cardoso for her contribution to this paper.

\section*{Annex}

As shown by the previous results, this method's performance depends on the initial guess, \textit{i.e.}, closer initial guesses take less time to converge to a good solution. The measurements can be obtained with measuring tape and anthropometric formulas but even if the doctor would need to guess, he would be able to do so with a very small range of error (depending on the experience). However, estimating the  breast biomechanical parameters is not as easy. So, while a doctor can estimate the measurements with an error up to $10\%$, we can assume that, even with some experience, a doctor would guess with an error that could go up to $50\%$.

With these assumptions we performed additional tests to evaluate the performance of the iterative estimator proposed in this paper. One of the tests concerns the average error obtained with the estimation of the parameters with an initial guess that differs the reference parameters up to $50\%$. We also show for these tests the error one obtains (in average) of the measurements, \textit{i.e.}, we evaluate the difference between the reference measurements and the estimated measurements. The second test evaluates the robustness of the method maintaining the initial guess configuration (differences up to $50\%$) and assuming that the errors of the doctors obtaining the measurements can go up to $10\%$. 

The values that are shown in the subsections below, are the average values of 50 cases using three different sized meshes (coarse, thin and ultra-thin).

\subsection*{Estimator performance in practical cases}

We show in table \ref{tab:nn:it:rob_50} the average error of estimating the breast parameters considering an initial guess which differs the reference parameters up to $50\%$. 

\begin{table}[ht]
\setlength{\tabcolsep}{1.5pt}
\renewcommand{\arraystretch}{1.25}
\centering
\begin{tabular}{|c|c|c|c|c|c|c|c|c|c|}
\hline
\textbf{\begin{tabular}[c]{@{}c@{}}Error per\\ Parameter\\ (\%)\end{tabular}} & \textit{\textbf{$R$}} & \textit{\textbf{$H$}} & \textit{\textbf{$\lambda$}} & \textit{\textbf{$\mu$}} & \textit{\textbf{$\lambda_{p}$}} & \textit{\textbf{$\mu_{p}$}} & \textit{\textbf{\begin{tabular}[c]{@{}c@{}}avg\\ total\end{tabular}}} & \textit{\textbf{cycles}} & \textit{\textbf{\begin{tabular}[c]{@{}c@{}}time\\ (s)\end{tabular}}} \\ \hline
\textbf{\begin{tabular}[c]{@{}c@{}}Coarse\\ (N=226)\end{tabular}} & 2.30 & 2.71 & 5.28 & 3.07 & 9.41 & 15.72 & 6.42 & 12 & 196 \\ \hline
\textbf{\begin{tabular}[c]{@{}c@{}}Thin\\ (N=1014)\end{tabular}} & 2.42 & 3.78 & 5.11 & 2.68 & 9.24 & 15.19 & 6.40 & 13 & 696 \\ \hline
\textbf{\begin{tabular}[c]{@{}c@{}}Ultra-Thin\\ (N=2498)\end{tabular}} & 2.56 & 3.83 & 5.07 & 2.48 & 9.13 & 14.43 & 6.25 & 14 & 1587 \\ \hline
\end{tabular}
\caption{Mean difference in percentage between the reference parameters and the estimated parameters using three mesh sizes: Coarse, Thin and Ultra-Thin. Time spent in average in seconds as well as the number of iterations (cycles) required by the iterative method to converge and produce a good result.}
\label{tab:nn:it:rob_50}
\end{table}

The values show a good performance by the iterative method with average errors rounding the $6\%$. The most problematic parameters are the skin mechanical parameters $\lambda_{p}$ and, specially, $\mu_{p}$ which confirms the assumptions that the measurements used to estimate these parameters are not ideal. It is also possible to see, that the error is better for thinner meshes. However, looking at every parameter it is possible to confirm that coarser meshes estimate the geometrical parameters more accurately while the thinner meshes estimate the mechanical parameters with less error. The sensitivity analysis showed that an error at a certain parameter is compensated by changing the other parameters. So it is possible that with more thinner meshes the error obtained with the geometrical parameters will be so high that will affect the estimation of the mechanical parameters. This suggests that it is possible that there is a mesh size configuration that will be optimal (balance between the error of the geometrical and the mechanical parameters).

These small values of error of the parameters don't mean anything if the error of the measurements obtained with the estimated parameters is high. So we present in table \ref{table::nn:test_measure} the average difference between the reference measurements $\overline{M}$ and the estimated measurements $M(\Lambda)$.

\begin{table}[ht]
\setlength{\tabcolsep}{1.5pt}
\renewcommand{\arraystretch}{1.5}
\centering
\begin{tabular}{|c|c|c|c|c|c|c|c|c|c|}
\hline
\textbf{Mesh} & \textit{\textbf{$e_{m_{1}}$}} & \textit{\textbf{$e_{m_{2}}$}} & \textit{\textbf{$e_{m_{3}}$}} & \textit{\textbf{$e_{m_{4}}$}} & \textit{\textbf{$e_{m_{5}}$}} & \textit{\textbf{$e_{m_{6}}$}} & \textit{\textbf{$e_{m_{7}}$}} & \textit{\textbf{$e_{m_{8}}$}} & \textit{\textbf{$e_{m_{9}}$}} \\ \hline
\textbf{Coarse} & 0.47 & 0.64 & 0.87 & 0.52 & 6.53 & 0.73 & 0.51 & 0.80 & 0.76 \\ \hline
\textbf{Medium} & 0.48 & 0.61 & 0.82 & 0.51 & 6.21 & 0.67 & 0.63 & 0.83 & 0.75 \\ \hline
\textbf{Thin} & 0.48 & 0.50 & 0.79 & 0.51 & 6.18 & 0.61 & 0.51 & 0.85 & 0.73 \\ \hline
\textbf{Mesh} & \textit{\textbf{$e_{m_{10}}$}} & \textit{\textbf{$e_{m_{11}}$}} & \textit{\textbf{$e_{m_{12}}$}} & \textit{\textbf{$e_{m_{13}}$}} & \textit{\textbf{$e_{m_{14}}$}} & \textit{\textbf{$e_{m_{15}}$}} & \textit{\textbf{\begin{tabular}[c]{@{}c@{}}avg\\ total\end{tabular}}} & \multicolumn{2}{c|}{\textit{\textbf{\begin{tabular}[c]{@{}c@{}}cost\\ function\end{tabular}}}} \\ \hline
\textbf{Coarse} & 8.23 & 0.67 & 0.43 & 0.51 & 0.56 & 7.52 & 1.98 & \multicolumn{2}{c|}{1.99e-6} \\ \hline
\textbf{Medium} & 7.11 & 0.73 & 0.52 & 0.57 & 0.55 & 5.87 & 1.79 & \multicolumn{2}{c|}{1.81e-6} \\ \hline
\textbf{Thin} & 7.17 & 0.67 & 0.47 & 0.52 & 0.55 & 5.73 & 1.75 & \multicolumn{2}{c|}{1.72e-6} \\ \hline
\end{tabular}
\caption{Mean value error in \% between the reference and estimated measurements for each mesh using the iterative estimator with considering an initial guess that differs the reference parameters up to $50\%$. Last lower column shows the mean value of the cost function.}
\label{table::nn:test_measure}
\end{table}

The values show average values of error inferior to $2\%$. In fact most measurement errors are inferior to $1\%$ with the exception of measurements $m_{5}$, $m_{10}$ and $m_{15}$. But these values are usually of a few millimetres which makes them negligible. These results are very important to prove this method reliability because they prove that this method is capable of estimating the breast parameters that incur in error's inferior to the millimetre.

\subsection*{Estimator robustness in practical cases}

We estimated the robustness of the method assuming its use by doctors. As mentioned previously we assume they can obtain the breast measurements with errors that can go up to $10\%$. 

\begin{table}[ht]
\setlength{\tabcolsep}{1.5pt}
\renewcommand{\arraystretch}{1.25}
\centering
\begin{tabular}{|c|c|c|c|c|c|c|c|}
\hline
\textbf{\begin{tabular}[c]{@{}c@{}}Error per\\ Parameter\\ (\%)\end{tabular}} & \textit{\textbf{$R$}} & \textit{\textbf{$H$}} & \textit{\textbf{$\lambda$}} & \textit{\textbf{$\mu$}} & \textit{\textbf{$\lambda_{p}$}} & \textit{\textbf{$\mu_{p}$}} & \textit{\textbf{\begin{tabular}[c]{@{}c@{}}avg\\ total\end{tabular}}} \\ \hline
\textbf{\begin{tabular}[c]{@{}c@{}}Coarse\\ (N=226)\end{tabular}} & 8.05 & 9.49 & 15.31 & 6.45 & 19.76 & 39.3 & 16.39 \\ \hline
\textbf{\begin{tabular}[c]{@{}c@{}}Thin\\ (N=1014)\end{tabular}} & 8.47 & 13.23 & 14.82 & 5.63 & 19.40 & 37.98 & 16.59 \\ \hline
\textbf{\begin{tabular}[c]{@{}c@{}}Ultra-Thin\\ (N=2498)\end{tabular}} & 8.96 & 13.41 & 14.70 & 5.21 & 19.17 & 36.08 & 16.25 \\ \hline
\end{tabular}
\caption{Mean value of the error in (\%) between the reference parameters and the approximation obtained by the IM with three different sized meshes considering errors in the measurements up to 10\%.}
\label{table::nn:nn_robust}
\end{table}

Table \ref{table::nn:nn_robust} shows the values of error on the estimated parameters with the inclusion of errors of the measurements. It is possible to see that in average the impact of the error is visible but not considerable, \textit{i.e.}, with errors in the measurements that can go up to $10\%$ we observe a n increase in the average error per parameter of also $10\%$ (average of error of $16\%$ compared to the average error per parameter of $6\%$ in table \ref{tab:nn:it:rob_50}). The most important aspect of these results is that they show that this method is robust and therefore is reliable.

\section*{Conflict of Interest}

The authors declare that there is no conflict of interest regarding the content of this article.


%
%

\bibliographystyle{spmpsci}      
\bibliography{refs.bib}   


\end{document}